\documentclass{article}
\usepackage{amsmath,amssymb,amsthm}
\usepackage{epic}

\newtheorem{theorem}{Theorem}[section]
\newtheorem{conjecture}[theorem]{Conjecture}
\newtheorem{corollary}[theorem]{Corollary}

\newtheorem{lemma}[theorem]{Lemma}
\theoremstyle{definition}

\theoremstyle{remark}

\newtheorem{example}[theorem]{Example}

\newcommand{\PRE}[2]{#1 \leq #2}
\newcommand{\SUF}[2]{#1 \preccurlyeq #2}
\newcommand{\eps}{\varepsilon}
\newcommand{\per}{\partial}

\title{Periodicity and Unbordered Words:\\
       A Proof of the Extended Duval Conjecture}

\author{TERO HARJU and
        DIRK NOWOTKA \\ Turku Centre for Computer Science
                        (TUCS), \\
                        Department of Mathematics,
                        University of Turku}
\date{May 2003}

\begin{document}

\maketitle

\begin{abstract}
  The relationship between the length of a word and
  the maximum length of its unbordered factors
  is investigated in this paper.
  Consider a finite word $w$ of length $n$.
  We call a word bordered, if it has a proper
  prefix which is also a suffix of that word.
  Let $\mu(w)$ denote the maximum length of all
  unbordered factors of~$w$, and let $\per(w)$ denote
  the period of~$w$. Clearly, $\mu(w)\leq\per(w)$.

  We establish that $\mu(w)=\per(w)$,
  if $w$ has an unbordered prefix of length
  $\mu(w)$ and \hbox{$n\geq 2\mu(w)-1$}. This bound
  is tight and solves the stronger version of a~21~years
  old conjecture by Duval. It follows from this result that,
  in general, $n\geq 3\mu(w)-2$ implies $\mu(w)=\per(w)$
  which gives an improved bound for the question
  asked by~Ehrenfeucht and~Silberger in~1979.
\end{abstract}

\section{Introduction}\label{sec:introduction}
Periodicity and borderedness are two properties
of words which are investigated in this paper.
These two concepts---periodicity and borderedness---are
fundamental and play a r\^ole (explicitly or
implicitly) in many areas. Just a few of those areas
are string searching algorithms
\cite{KMP:77,BoyerMoore:77,CrochemorePerrin:91},
data compression \cite{ZivLempel:77,CMRS:99}, and
codes \cite{BerstelPerrin:85}, which are classical
examples, but also computational biology, e.g.,
sequence assembly~\cite{MaSk:95} or~superstrings~\cite{BJJ:97},
and serial data communications systems \cite{BylanskiIngram:80}
are areas among others where periodicity and borderedness of words
(sequences) are important concepts.
It is well known that these two word properties do not
exist independently from each other. However, it is somewhat
surprising that no clear relation has been established so far,
despite the fact that this basic question has been around for
more than 20~years.

Let us consider a finite word (a sequence of~letters) $w$.
We denote the length of~$w$ by~$|w|$ and call a subsequence
of consecutive letters of a~word \emph{factor}. The period
of $w$, denoted by~$\per(w)$, is the smallest positive
integer $p$ such that the $i$-th letter equals the $(i+p)$-th
letter for all $1\leq i\leq |w|-p$. Let $\mu(w)$ denote
the maximum length of all unbordered factors of~$w$. A~word is
bordered, if it has a~proper prefix that is also a~suffix,
where we call a prefix proper, if it is neither empty nor
contains the entire word.
For the investigation of the relationship between $|w|$
and the maximality of $\mu(w)$, that is, $\mu(w)=\per(w)$,
we consider the special case where the longest unbordered
prefix of a~word is of the maximum length, that is,
no unbordered factor is longer than that prefix.
Let $w$ be an unbordered word. Then a~word $wu$
is a~\emph{Duval extension} (of~$w$), if every
unbordered factor of~$wu$ has at most length $|w|$,
that is, $\mu(wu)=|w|$. We call $wu$ \emph{trivial}
Duval extension, if $\per(wu)=|w|$, or with other words,
if $u$ is a~prefix of $w^k$ for some $k\geq 1$.
For example, let $w=abaabb$ and $u=aaba$. Then
$wu=abaabbaaba$ is a~nontrivial Duval extension
of~$w$ since (\textit{i}) $w$ is unbordered,
(\textit{ii}) all factors of~$wu$ longer than $w$
are bordered, that is, $|w|=\mu(wu)=6$,
and (\textit{iii}) the period
of~$wu$ is~$7$, and hence, $\per(wu)>|w|$.
Note, that this example satisfies $|u|=|w|-2$.

In 1979 a~line of research was initiated
\cite{EhrSil:79,AssPou:79,Duval:82}
exploring the relationship between the length
of a~word~$w$ and~$\mu(w)$.
In~1982 these efforts culminated in the following
result by~Duval: If $|w|\geq4\mu(w)-6$ then \hbox{$\per(w)=\mu(w)$}.
However, it was conjectured \cite{AssPou:79}
that $|w|\geq3\mu(w)$ implies
$\per(w)=\mu(w)$ which follows if Duval's conjecture
\cite{Duval:82} holds true.
\begin{conjecture}\label{conj:1}
  Let $wu$ be a nontrivial Duval extension of~$w$.
  Then $|u|<|w|$.
\end{conjecture}
After that, no progress was recorded, to the best
of our knowledge, for 20~years. However, the topic
remained popular, see for example Chapter~8 in~\cite{Loth:2}.
The most recent results are by Mignosi
and Zamboni~\cite{MiZa:02} and
the authors of this article~\cite{HaNo:02}.
However, not Duval's conjecture but rather its opposite
is investigated in those papers, that is: Which words admit only
trivial Duval extensions? It is shown \cite{MiZa:02}
that unbordered, finite factors of Sturmian words allow
only trivial Duval extensions, with other words, if an unbordered,
finite factor of a Sturmian word of length $\mu(w)$ is a~prefix
of $w$, then $\per(w)=\mu(w)$. Sturmian words are binary infinite words
of minimal subword complexity, that is, a Sturmian word contains
exactly $n+1$ different factors of length~$n$ for every $n\geq 1$;
see \cite{MoHe:40} or Chapter~2 in~\cite{Loth:2}.
That result was later improved \cite{HaNo:02} by showing that
Lyndon words~\cite{Lyndon:54} allow only trivial Duval
extensions and the fact that every unbordered, finite factor
of a Sturmian word is a Lyndon word. A~Lyndon word is a word
that is minimal among all its conjugates with respect to some
lexicographic order, where a word $uv$ is a conjugate of $vu$.

The main result in this paper is a proof of
the extended version of Conjecture~\ref{conj:1}.
\newcounter{maintheorem}
\setcounter{maintheorem}{\value{theorem}}
\newcounter{secintro}
\setcounter{secintro}{\value{section}}
\begin{theorem}\label{thm:main}
  Let $wu$ be a Duval nontrivial extension of~$w$.
  Then $|u|<|w|-1$.
\end{theorem}
The example mentioned above shows that this bound
on the length of a~nontrivial Duval extension
is tight.
Theorem~\ref{thm:main} implies the truth
of Duval's conjecture, as well as,
the following corollary (for any word $w$).
\newcounter{maincorollary}
\setcounter{maincorollary}{\value{theorem}}
\newcounter{secintroo}
\setcounter{secintroo}{\value{section}}
\begin{corollary}\label{cor:1}
  If $|w|\geq3\mu(w)-2$, then \hbox{$\per(w)=\mu(w)$}.
\end{corollary}
This corollary confirms the conjecture by Assous
and~Pouzet in~\cite{AssPou:79} about a~question asked
by Ehrenfeucht and~Silberger in~\cite{EhrSil:79}.

Our main result, Theorem~\ref{thm:main}, is presented in
Section~\ref{sec:main}, which uses the notations
introduced in Section~\ref{sec:notations} and preliminary results
from Section~\ref{sec:lemmas}. We conclude with
Section~\ref{sec:concl}.

\section{Notations}\label{sec:notations}
In this section we introduce the notations of this paper. We refer
to~\cite{Loth:1,Loth:2} for more basic and general definitions.

We consider a finite alphabet $A$ of letters. Let $A^\ast$
denote the monoid of~all finite words over $A$ including the
empty word, denoted by~$\eps$.
Let \hbox{$w=w_{(1)}w_{(2)}\cdots w_{(n)}$} where $w_{(i)}$
is a~letter, for every $1\leq i\leq n$. We denote
the length $n$ of~$w$ by~$|w|$. An~integer $1\leq p\leq n$
is a~\emph{period} of~$w$, if $w_{(i)}=w_{(i+p)}$ for all
$1\leq i\leq n-p$. The smallest period of $w$ is called
the \emph{minimum period} (or simply, the period) of~$w$,
denoted by~$\per(w)$.
A~nonempty word $u$ is called a~\emph{border} of a~word~$w$,
if $w=uv=v'u$ for some suitable words $v$ and~$v'$.
We call $w$ \emph{bordered}, if it has a~border that is shorter
than $w$, otherwise $w$ is called \emph{unbordered}.
Note, that every bordered word $w$ has a~minimum border~$u$
such that $w=uvu$, where $u$ is unbordered.
Let $\mu(w)$ denote the maximum length of unbordered factors of~$w$.
Suppose $w=uv$, then $u$ is called a~\emph{prefix} of $w$, denoted
by $\PRE{u}{w}$, and $v$ is called a~\emph{suffix} of~$w$,
denoted by $\SUF{v}{w}$.
Let $u,v\neq\eps$. Then we say that $u$ \emph{overlaps $v$
from the left} or \emph{from the right}, if there is a word
$w$ such that $|w|<|u|+|v|$, and $\PRE{u}{w}$ and $\SUF{v}{w}$,
or $\PRE{v}{w}$ and $\SUF{u}{w}$, respectively.
We say that $u$ \emph{overlaps} (intersects) with $v$,
if either $v$ is a factor of~$u$ or $u$ is a factor of~$v$
or $u$ overlaps $v$ from the left or~right.

Let us consider the following examples.
Let \hbox{$A=\{a, b\}$} and $u, v, w\in A^\ast$
such that $u=abaa$ and $v=baaba$ and $w=abaaba$.
Then $|w|=6$, and $3$, $5$, and $6$ are periods
of $w$, and $\per(w)=3$. We have that $a$ is
the shortest border of $u$ and $w$, whereas
$ba$ is the shortest border of $v$.
We have $\mu(w)=3$. We also have that
$u$ and $v$ overlap since $\PRE{u}{w}$
and $\SUF{v}{w}$ and $|w|<|u|+|v|$.

We continue with some more notations.
Let $w$ and $u$ be nonempty words
where $w$ is also unbordered.
We call $wu$ a~\emph{Duval extension} of~$w$,
if every factor of $wu$ longer than $|w|$ is bordered,
that is, $\mu(wu)=|w|$.
A Duval extension $wu$ of $w$ is called \emph{trivial},
if $\per(wu)=\mu(wu)=|w|$.
A~nontrivial Duval extension $wu$ of $w$
is called \emph{minimal}, if $u$ is of minimal length,
that is, \hbox{$u=u'a$} and $w=u'bw'$ where $a, b\in A$
and $a\neq b$.
\begin{example}
  Let $w=abaabbabaababb$ and \hbox{$u=aaba$}. Then
  \[
    w.u = abaabbabaababb.aaba
  \]
  (for the sake of readability,
  we use a dot to mark where $w$ ends)
  is a nontrivial Duval extension of~$w$ 
  of length $|wu|=18$, where
  \hbox{$\mu(wu)=|w|=14$} and \hbox{$\per(wu)=15$}.
  However, $wu$ is not a minimal Duval extension,
  whereas
  \[
    w.u' = abaabbabaababb.aa
  \]
  is minimal, with $\PRE{u'=aa}{u}$.
  Note, that $wu$ is not the longest nontrivial
  Duval extension of $w$ since
  \[
    w.v = abaabbabaababb.abaaba
  \]
  is longer, with $v=abaaba$ and $|wv|=20$ and
  $\per(wv)=17$. One can check that $wv$
  is a nontrivial Duval extension of~$w$ of maximum
  length, and at the same time $wv$ is also
  a minimal Duval extension of $w$.
\end{example}

Let an integer~$p$ with \hbox{$1\leq p < |w|$} be called
\emph{point} in~$w$. Intuitively,
a~point~$p$ denotes the place between $w_{(p)}$
and~$w_{(p+1)}$ in~$w$. A~nonempty word $u$ is called
a~\emph{repetition word} at point~$p$ if~$w=xy$ with
$|x|=p$ and there exist $x'$ and~$y'$ such
that $\SUF{u}{x'x}$ and \hbox{$\PRE{u}{yy'}$}.
For a~point~$p$ in~$w$, let
\[
  \per(w, p)=\min\bigl\{|u| \bigm| u
    \text{ is a repetition word at } p\bigr\}
\]
denote the \emph{local period} at point~$p$ in~$w$.
Note, that the repetition word of length $\per(w, p)$ at
point~$p$ is necessarily unbordered
and \hbox{$\per(w, p)\leq\per(w)$}. A~factorization \hbox{$w=uv$},
with $u, v\neq\eps$ and~$|u|=p$, is called \emph{critical},
if $\per(w, p)=\per(w)$, and, if this holds, then $p$
is called \emph{critical point}.

\begin{example}
  The word
  \[
    w = ab.aa.b
  \]
  has the period $\per(w)=3$ and two critical points,
  $2$ and $4$, marked by dots. The shortest repetition words
  at the critical points are $aab$ and $baa$, respectively.
  Note, that the shortest repetition words at the remaining
  points $1$ and $3$ are $ba$ and $a$, respectively.
\end{example}

\section{Preliminary Results}\label{sec:lemmas}
We state some auxiliary and well-known results
about repetitions and borders in
this section which will be used to prove
Theorem~\ref{thm:main}, in Section~\ref{sec:main}.

\begin{lemma}\label{lem:3}
  Let $zf=gzh$ where $f, g\neq\eps$. Let $az'$
  be the maximum unbordered prefix of $az$.
  If $az$ does not occur in $zf$,
  then $agz'$ is unbordered.
\end{lemma}
\begin{proof}
  Assume $agz'$ is bordered, and let $y$ be its
  shortest border. In particular, $y$ is unbordered.
  If $|z'|\geq|y|$ then $y$ is a border of $az'$
  which is a~contradiction.
  If $|az'|=|y|$ or $|az|<|y|$ then $az$ occurs
  in $zf$ which is again a contradiction.
  If \hbox{$|az'|<|y|\leq|az|$} then $az'$
  is not maximum
  since $y$ is unbordered; a~contradiction.
\end{proof}

The proof of the following lemma is easy.

\begin{lemma}\label{lem:5}
  Let $w$ be an unbordered word and $\PRE{u}{w}$
  and $\SUF{v}{w}$. Then $uw$ and $wv$ are unbordered.
\end{lemma}

The critical factorization theorem is one of the main
results about periodicity of words. A weak version
of it was first conjectured by Sch\"utzenberger~\cite{Sch:76}
and proved by C\'esari and {Vin\-cent}~\cite{CesariVincent:78}.
It was developed into its current form by~Duval~\cite{Duval:79}.
We refer to~\cite{HaNo:02b} for a~short proof of the CFT.
\begin{theorem}[CFT]\label{thm:cft}
  Every word $w$, with \hbox{$|w|\geq 2$}, has at least
  one critical factorization $w=uv$, with $u, v\neq\eps$
  and $|u|<\per(w)$, i.e., $\per(w, |u|) = \per(w)$.
\end{theorem}

We have the following two lemmas about properties
of critical factorizations.

\begin{lemma}\label{lem:6}
  Let $w=uv$ be unbordered and $|u|$ be a critical point
  of~$w$. Then $u$ and $v$ do not overlap.
\end{lemma}
\begin{proof}
  Note, that $\per(w,|u|)=\per(w)=|w|$ since $w$ is unbordered.
  Let $|u|\leq|v|$ without restriction of generality.
  Assume that $u$ and $v$ overlap.
  If $u=u's$ and $v=sv'$, then $\per(w,|u|)\leq|s|<|w|$.
  On the other hand, if $u=su'$ and $v=v's$,
  then $w$ is bordered with $s$.
  Finally, if $v=sut$ then $\per(w,|u|)\leq|su|<|w|$.
\end{proof}

The next result follows directly from Lemma~\ref{lem:6}.
\begin{lemma}\label{lem:4}
  Let $u_0u_1$ be unbordered and $|u_0|$ be a critical point
  of~$u_0u_1$. Then for any word $x$, we have $u_ixu_{i+1}$,
  where the indices are modulo~$2$, is either unbordered or has
  a minimum border $g$ such that $|g|\geq|u_0|+|u_1|$.
\end{lemma}

The next theorem states a basic fact about minimal
Duval extensions. See~\cite{HaNo:03} for a proof of it.

\begin{theorem}\label{thm:1}
  Let $wu$ be a minimal Duval extension of~$w$.
  Then $u$ occurs in~$w$.
\end{theorem}

The following Lemmas~\ref{lem:7}, \ref{lem:2}
and~\ref{lem:1} and Corollary~\ref{cor:1} are
given in~\cite{Duval:82}.
Let $a_0, a_1\in A$, with $a_0\neq a_1$,
and $t_0\in A^\ast$. Let the sequences
$(a_i)$, $(s_i)$, $(s'_i)$, $(s''_i)$, and $(t_i)$,
for $i\geq 1$, be defined by
\begin{itemize}
  \item $a_i = a_{i\ (\mathrm{mod}\ 2)}$, that is,
    $a_i = a_0$ or $a_i = a_1$, if $i$ is even
    or odd, respectively,
  \item $s_i$ such that $a_i s_i$ is the
    shortest border of~$a_i t_{i-1}$,
  \item $s'_i$ such that $a_{i+1} s'_i$
    is the longest unbordered prefix of $a_{i+1}s_i$,
  \item $s''_i$ such that $s'_i s''_i=s_i$,
  \item $t_i$ such that $t_i s''_i = t_{i-1}$.
\end{itemize}

For any parameters of the above definition,
the following holds.

\begin{lemma}\label{lem:7}
  For any $a_0$, $a_1$, and $t_0$ there exists
  an $m\geq 1$ such that
  \[
    |s_1|<\cdots<|s_m|=|t_{m-1}|\leq\cdots\leq|t_0|
  \]
  and $s_m=t_{m-1}$ and $|t_0|\leq|s_m|+|s_{m-1}|$.
\end{lemma}

\begin{lemma}\label{lem:2}
  Let $\PRE{z}{t_0}$ such that $a_0z$ and $a_1z$
  do not occur in $t_0$. Let $a_0z_0$ and $a_1z_1$
  be the longest unbordered prefixes of $a_0z$
  and $a_1z$, respectively. Then
  \begin{enumerate}
    \item if $m=1$ then $a_0t_0$ is unbordered,
    \item if $m>1$ is odd, then $a_1s_m$ is
      unbordered and $|t_0|\leq|s_m|+|z_0|$,
    \item if $m>1$ is even, then $a_0s_m$ is
      unbordered and $|t_0|\leq|s_m|+|z_1|$.
  \end{enumerate}
\end{lemma}

\begin{lemma}\label{lem:1}
  Let $v$ be an unbordered factor of~$w$ of length
  $\mu(w)$.
  If $v$ occurs twice in $w$, then $\mu(w)=\per(w)$.
\end{lemma}

\begin{corollary}
  Let $wu$ be a Duval extension of~$w$.
  If $w$ occurs twice in $wu$, then $wu$
  is a~trivial Duval extension.
\end{corollary}

\section{Main Result}\label{sec:main}
The extended Duval conjecture is proven in this section.
\newcounter{EGAL}
\setcounter{EGAL}{\value{theorem}}
\setcounter{theorem}{\value{maintheorem}}
\newcounter{EGAL:sec}
\setcounter{EGAL:sec}{\value{section}}
\setcounter{section}{\value{secintro}}
\begin{theorem}
  Let $wu$ be a nontrivial Duval extension
  of~$w$. Then $|u|<|w|-1$.
\end{theorem}
\setcounter{section}{\value{EGAL:sec}}
\setcounter{theorem}{\value{EGAL}}
\begin{proof}
  Recall that every factor of $wu$ which is longer
  than $|w|$ is bordered since $wu$ is
  a~Duval extension of~$w$.
  Let $z$ be the longest suffix of $w$ that occurs
  twice in $zu$.

  If $z=\eps$ then
  $\SUF{a}{w}$ and $u=b^j$,
  where $a, b\in A$ and $a\neq b$ and $j\geq 1$,
  but now $|u|<|w|$ since $ab^j$ is unbordered.
  Moreover, $w=b^k a w' a$ with \hbox{$k<j$}, otherwise
  $wu$ is a trivial Duval extension, and either
  $aw'ab^j$ is bordered, in this case it follows $j\leq|w'|$,
  or $aw'ab^j$ is unbordered. In both cases it follows
  $|u|<|w|-1$.

  So, assume $z\neq\eps$.
  We have $z\neq w$ since $wu$ is otherwise trivial
  by Corollary~\ref{cor:1}.
  Let $a, b\in A$ be such that
  \[
    w = w' az \qquad\text{and}\qquad
    u = u' bz r
  \]
  and $z$ occurs in $zr$ only once, that is, $bz$
  matches the rightmost occurrence of~$z$ in~$u$.
  Note, that $bz$ does not overlap $az$ from the right,
  by Lemma~\ref{lem:5}, and therefore $u'$ exists,
  although it might be empty.
  Naturally, $a\neq b$ by the maximality of~$z$,
  and $w'\neq\eps$, otherwise \hbox{$\PRE{azu'bz}{wu}$} has
  either no border or $w$ is bordered (if $azu'bz$
  has a border not longer than $z$) or $az$ occurs
  in $zu$ (if $azu'bz$ has a border longer than~$z$);
  a~contradiction in any case.

  Let $az_0$ and $bz_1$ denote the longest unbordered
  prefix of $az$ and $bz$, respectively.
  Let $a_0=a$ and $a_1=b$ and $t_0=zr$ and
  the integer $m$ be defined as in Lemma~\ref{lem:2}.
  We have then a word $s_m$, with its properties
  defined by Lemma~\ref{lem:2}, such that
  \[
    t_0=s_m t' \ .
  \]
  Consider $azu'bz_0$. We have that $az$ and $azu'bz_0$
  are both prefixes of $a_0zu$, and $bz_0$ is
  a suffix of $azu'bz_0$ and $az$ does not occur in
  $zu'bz_0$. It follows from Lemma~\ref{lem:3} that
  $azu'bz_0$ is unbordered, and hence,
  \begin{equation}
    |azu'bz_0|\leq|w| \ . \label{eq:1}
  \end{equation}
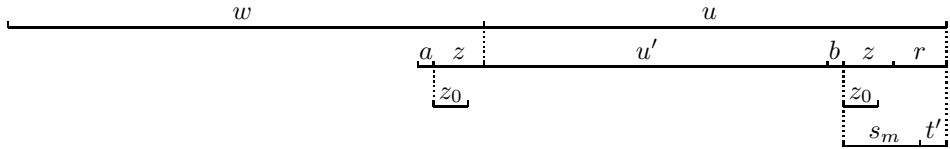
  \begin{center}
  \begin{picture}(355,55)(0,0)
    \put(85,48){\makebox{$w$}}
    \put(262.5,48){\makebox{$u$}}
    \drawline(0,47)(0,45)(180,45)(180,47)
    \drawline(180,47)(180,45)(355,45)(355,47)

    \put(155.5,33){\makebox{$a$}}
    \put(168,33){\makebox{$z$}}
    \drawline(155,32)(155,30)(161,30)(161,32)(161,30)
         (180,30)(180,32)
    \dottedline{2}(180,32)(180,45)
    \put(310.6,33){\makebox{$b$}}
    \put(323,33){\makebox{$z$}}
    \drawline(310,32)(310,30)(316,30)(316,32)(316,30)
         (335,30)(335,32)
    \put(237.5,33){\makebox{$u'$}}
    \drawline(180,32)(180,30)(310,30)(310,32)
    \put(342.5,33){\makebox{$r$}}
    \drawline(335,32)(335,30)(355,30)(355,32)
    \dottedline{2}(355,32)(355,45)

    \put(163,18){\makebox{$z_0$}}
    \drawline(161,17)(161,15)(174,15)(174,17)
    \dottedline{2}(161,17)(161,30)
    \put(318,18){\makebox{$z_0$}}
    \drawline(316,17)(316,15)(329,15)(329,17)
    \dottedline{2}(316,17)(316,30)

    \put(325.5,3){\makebox{$s_m$}}
    \put(347,3){\makebox{$t'$}}
    \drawline(316,2)(316,0)(345,0)(345,2)(345,0)
         (355,0)(355,2)
    \dottedline{2}(355,2)(355,30)
    \dottedline{2}(316,2)(316,15)
  \end{picture}
  \end{center}

  \textbf{Case:} Suppose that $m$ is even. Then we have
  $2\leq m$ and $as_m$
  ($= a_m s_m$) is unbordered and \hbox{$|t_0|\leq|s_m|+|z_1|$}
  by Lemma~\ref{lem:2}.

  Suppose $|t_0|=|s_m|+|z_1|$ and $z_1=z$.
  Then \hbox{$|s_{m-1}|=|z|$} by Lemma~\ref{lem:7}.
  Note, that $\PRE{\PRE{s_i}{t_{i-1}}}{t_0}$ for all $1\leq i\leq m$,
  and hence, it follows that $\PRE{s_i}{z}$ for all $1\leq i<m$.
  In particular, $s_{m-1}=z$.
  We have that $bz$ ($= a_1s_{m-1}$)
  is a~border of $bt_{m-2}$ ($= a_1 t_{m-2}$).
  But now, $bz$ occurs in $t_0$, and hence, in $u$,
  since $\PRE{t_i}{t_0}$, for all $0\leq i<m$,
  which is a contradiction.

  So, assume that $|t_0|<|s_m|+|z_1|$ or $|z_1|<|z|$.
  Suppose $|s_m|\leq|z_0|$. Then $|azu'bz_0|\leq|w|$ and 
  \begin{align*}
    |u| &= |azu| - |z| - 1 \\
        &= |azu'bz_0|-|z_0| + |t_0| - |z| - 1 \\
        &< |azu'bz_0|-|z_0| + |s_m| + |z_1| - |z| - 1 \\
        &\leq |w| + |z_1| - |z| - 1 \\
        &\leq |w| - 1
  \end{align*}
  if $|t_0|<|s_m|+|z_1|$, or
  \begin{align*}
    |u| &= |azu| - |z| - 1 \\
        &= |azu'bz_0|-|z_0| + |t_0| - |z| - 1 \\
        &\leq |azu'bz_0|-|z_0| + |s_m| + |z_1| -  |z| - 1 \\
        &\leq |w| + |z_1| - |z| - 1 \\
        &< |w| - 1
  \end{align*}
  if $|z_1|<|z|$. We have $|u|<|w|-1$ in both cases.

  Let then $|s_m|>|z_0|$. We have that $as_m$ is
  unbordered, and since $az_0$ is the longest unbordered
  prefix of $az$, we have $\PRE{az}{as_m}$, and hence,
  $|z|\leq|s_m|$. Now, $azu'bs_m$ is unbordered
  otherwise its shortest border is longer than $az$,
  since no prefix of $az$ is a suffix
  of $as_m$, and $az$ occurs in $u$; a~contradiction.
  So, \hbox{$|azu'bs_m|\leq |w|$} and $|u|<|w|-1$, since either
  $|z_1|\leq|z|$ or $|t_0|<|s_m|+|z_1|$.

  \textbf{Case:} Suppose that $m$ is odd. Then $bs_m$
  \hbox{($= a_m s_m$)} is an unbordered word and
  \hbox{$|t_0|\leq |s_m|+|z_0|$};
  see Lemma~\ref{lem:2}. Surely $s_m\neq\eps$.

  If $|s_m|<|z|$, then $|u|<|w|-1$ since
  \[
    |u|=|azu'bz_0|-|bz_0|+|bt_0|-|az|
  \]
  and $|azu'bz_0|\leq|w|$, by~\eqref{eq:1},
  and $|t_0|\leq |s_m|+|z_0|$.

  Assume thus that $|s_m|\geq|z|$, and hence, also $\PRE{z}{s_m}$.
  Since $s_m\neq\eps$, we have $|bs_m|\geq 2$, and therefore,
  by the critical factorization theorem, there exists
  a~critical point $p$ in $bs_m$ such that $bs_m = v_0 v_1$,
  where $|v_0|=p$.
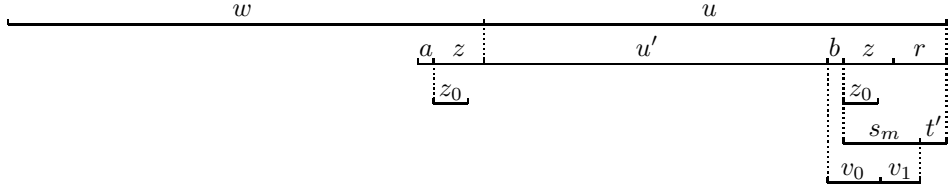
  \begin{center}
  \begin{picture}(355,70)(0,0)
    \put(85,63){\makebox{$w$}}
    \put(262.5,63){\makebox{$u$}}
    \drawline(0,62)(0,60)(180,60)(180,62)
    \drawline(180,62)(180,60)(355,60)(355,62)

    \put(155.5,48){\makebox{$a$}}
    \put(168,48){\makebox{$z$}}
    \drawline(155,47)(155,45)(161,45)(161,47)(161,45)
         (180,45)(180,47)
    \dottedline{2}(180,47)(180,60)
    \put(310.6,48){\makebox{$b$}}
    \put(323,48){\makebox{$z$}}
    \drawline(310,47)(310,45)(316,45)(316,47)(316,45)
         (335,45)(335,47)
    \put(237.5,48){\makebox{$u'$}}
    \drawline(180,47)(180,45)(310,45)(310,47)
    \put(342.5,48){\makebox{$r$}}
    \drawline(335,47)(335,45)(355,45)(355,47)
    \dottedline{2}(355,47)(355,60)

    \put(163,33){\makebox{$z_0$}}
    \drawline(161,32)(161,30)(174,30)(174,32)
    \dottedline{2}(161,32)(161,45)
    \put(318,33){\makebox{$z_0$}}
    \drawline(316,32)(316,30)(329,30)(329,32)
    \dottedline{2}(316,32)(316,45)

    \put(325.5,18){\makebox{$s_m$}}
    \put(347,18){\makebox{$t'$}}
    \drawline(316,17)(316,15)(345,15)(345,17)(345,15)
         (355,15)(355,17)
    \dottedline{2}(316,17)(316,30)
    \dottedline{2}(355,17)(355,45)

    \put(315,3){\makebox{$v_0$}}
    \put(333,3){\makebox{$v_1$}}
    \drawline(310,2)(310,0)
        (330,0)(330,2)(330,0)
        (345,0)(345,2)
    \dottedline{2}(310,2)(310,45)
    \dottedline{2}(345,2)(345,15)
  \end{picture}
  \end{center}
  In particular,
  \begin{equation}
    \PRE{bz}{v_0v_1} \ . \label{eq:13}
  \end{equation}
  Note, that if $s_m = z$ then $|z_0|<|z|$
  since $\SUF{b}{z_0}$ and $bs_m$ does not end with~$b$
  because it is unbordered. We have therefore
  in all cases
  \begin{equation}
    |z_0| < |v_0v_1| - 1 \ .  \label{eq:12}
  \end{equation}
  Let
  \[
    u = u'_0 v_0v_1 u_1
  \]
  be such that $v_0v_1$ does not occur in $u'_0$.
  Note, that $v_0v_1$ does not overlap with itself since
  it is unbordered, and $v_0$ and $v_1$ do not overlap
  by Lemma~\ref{lem:6}. Consider the prefix $wu'_0bz$
  of~$wu$ which is bordered and has a shortest
  border $g$ longer than $z$, and hence, $\SUF{bz}{g}$,
  otherwise $w$ is bordered since $\SUF{z}{w}$.
  Moreover, $\PRE{g}{w}$, for otherwise $az$ would occur in $u$,
  and hence, $bz$ occurs in~$w$. Let
  \[
    w = w_0 bz w_1
  \]
  such that $bz$ occurs in $w_0bz$ only once, that is, we
  consider the leftmost occurrence of $bz$ in $w$.
  Note, that
  \begin{equation}
    |w_0bz|\leq|g|\leq|u'_0bz| \label{eq:11}
  \end{equation}
  where the first inequality comes from the definition
  of~$w_0$ above and the second inequality from the
  fact that $|u'_0bz|<|g|$ implies that $w$ is bordered.
  Let
  \[
    f = bz w_1 u'_0 v_0v_1 \ .
  \]
  If $f$ is unbordered, then $|f|\leq|w|$, and hence,
  $|u'_0v_0v_1|\leq|w_0|$. Now, we have $|u'_0|<|w_0|$
  which contradicts~(\ref{eq:11}).

  Therefore, $f$ is bordered. Let $h$ be its
  shortest border.
  
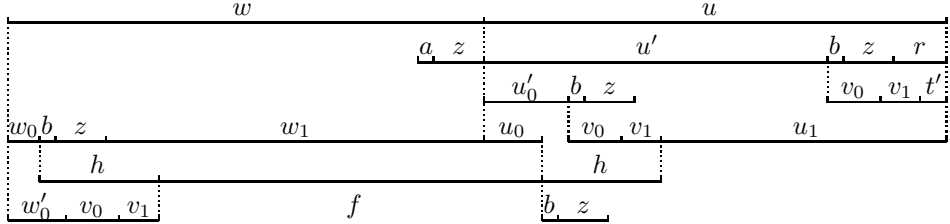
\begin{center}
  \begin{picture}(355,85)(0,0)
    \put(85,78){\makebox{$w$}}
    \put(262.5,78){\makebox{$u$}}
    \drawline(0,77)(0,75)(180,75)(180,77)
    \drawline(180,77)(180,75)(355,75)(355,77)

    \put(155.5,63){\makebox{$a$}}
    \put(168,63){\makebox{$z$}}
    \drawline(155,62)(155,60)(161,60)(161,62)(161,60)
         (180,60)(180,62)
    \dottedline{2}(180,62)(180,75)
    \put(310.6,63){\makebox{$b$}}
    \put(323,63){\makebox{$z$}}
    \drawline(310,62)(310,60)(316,60)(316,62)(316,60)
         (335,60)(335,62)
    \put(237.5,63){\makebox{$u'$}}
    \drawline(180,62)(180,60)(310,60)(310,62)
    \put(342.5,63){\makebox{$r$}}
    \drawline(335,62)(335,60)(355,60)(355,62)
    \dottedline{2}(355,62)(355,75)

    \put(190.5,48){\makebox{$u'_0$}}
    \put(212.6,48){\makebox{$b$}}
    \put(225,48){\makebox{$z$}}
    \drawline(180,47)(180,45)
        (212,45)(212,47)(212,45)
        (218,45)(218,47)(218,45)
        (237,45)(237,47)

    \put(315,48){\makebox{$v_0$}}
    \put(333,48){\makebox{$v_1$}}
    \drawline(310,47)(310,45)
        (330,45)(330,47)(330,45)
        (345,45)(345,47)
    \dottedline{2}(310,47)(310,60)

    \put(347,48){\makebox{$t'$}}
    \drawline(345,47)(345,45)
         (355,45)(355,47)
    \dottedline{2}(355,47)(355,60)

    \put(0,33){\makebox{$w_0$}}
    \put(12.6,33){\makebox{$b$}}
    \put(25,33){\makebox{$z$}}
    \put(103,33){\makebox{$w_1$}}
    \drawline(0,32)(0,30)
        (12,30)(12,32)(12,30)
        (18,30)(18,32)(18,30)
        (37,30)(37,32)(37,30)
        (180,30)(180,32)
    \dottedline{2}(0,32)(0,75)
    \dottedline{2}(180,32)(180,60)

    \put(185.5,33){\makebox{$u_0$}}
    \drawline(180,32)(180,30)(202,30)(202,32)

    \put(217,33){\makebox{$v_0$}}
    \put(235,33){\makebox{$v_1$}}
    \put(297,33){\makebox{$u_1$}}
    \drawline(212,32)(212,30)
        (232,30)(232,32)(232,30)
        (247,30)(247,32)(247,30)
        (355,30)(355,32)
    \dottedline{2}(212,32)(212,45)
    \dottedline{2}(355,32)(355,45)

    \put(31,18){\makebox{$h$}}
    \put(221,18){\makebox{$h$}}
    \drawline(12,17)(12,15)
        (57,15)(57,17)(57,15)
        (202,15)(202,17)(202,15)
        (247,15)(247,17)
    \dottedline{2}(12,17)(12,30)
    \dottedline{2}(202,17)(202,30)
    \dottedline{2}(247,17)(247,30)

    \put(5,3){\makebox{$w'_0$}}
    \put(27,3){\makebox{$v_0$}}
    \put(45,3){\makebox{$v_1$}}
    \drawline(0,2)(0,0)
        (22,0)(22,2)(22,0)
        (42,0)(42,2)(42,0)
        (57,0)(57,2)
    \dottedline{2}(0,2)(0,30)
    \dottedline{2}(57,2)(57,15)

    \put(202.6,3){\makebox{$b$}}
    \put(215,3){\makebox{$z$}}
    \drawline(202,2)(202,0)
        (208,0)(208,2)(208,0)
        (227,0)(227,2)
    \dottedline{2}(202,2)(202,15)

    \put(128,3){\makebox{$f$}}
  \end{picture}
  \end{center}
  Surely, $|bz|<|h|$ otherwise $v_0v_1$ is bordered
  by~\eqref{eq:13}. So, $\PRE{bz}{h}$. Moreover,
  \hbox{$|v_0v_1|\leq|h|$} otherwise $bz$ occurs in $s_m$
  contradicting our assumption that $bzr$ marks the
  rightmost occurrence of~$bz$ in~$u$. So, $\SUF{v_0v_1}{h}$,
  and $v_0v_1$ occurs in~$w$ since $\PRE{w_0h}{w}$
  by~\eqref{eq:11}.
  Let
  \[
    w_0 bz v' = w_0 h = w_0' v_0v_1 \ .
  \]
  Note, that $v_0v_1$ does not occur in $w'_0$ otherwise
  it occurs in $u'_0$ contradicting our assumption on~$u'_0$.
  Moreover, we have $\SUF{h=bzv'}{u'_0v_0v_1}$. Let
  \hbox{$u'_0v_0v_1=u_0h$}. Consider
  \[
    f_0 = w u_0 bz
  \]
  which has a shortest border $h_0$.
  \begin{center}
  \begin{picture}(355,85)(0,0)
    \put(85,78){\makebox{$w$}}
    \put(262.5,78){\makebox{$u$}}
    \drawline(0,77)(0,75)(180,75)(180,77)
    \drawline(180,77)(180,75)(355,75)(355,77)

    \put(155.5,63){\makebox{$a$}}
    \put(168,63){\makebox{$z$}}
    \drawline(155,62)(155,60)(161,60)(161,62)(161,60)
         (180,60)(180,62)
    \dottedline{2}(180,62)(180,75)
    \put(310.6,63){\makebox{$b$}}
    \put(323,63){\makebox{$z$}}
    \drawline(310,62)(310,60)(316,60)(316,62)(316,60)
         (335,60)(335,62)
    \put(342.5,63){\makebox{$r$}}
    \drawline(335,62)(335,60)(355,60)(355,62)
    \dottedline{2}(355,62)(355,75)

    \put(190.5,63){\makebox{$u'_0$}}
    \put(212.6,63){\makebox{$b$}}
    \put(225,63){\makebox{$z$}}
    \drawline(180,62)(180,60)
        (212,60)(212,62)(212,60)
        (218,60)(218,62)(218,60)
        (237,60)(237,62)

    \put(217,48){\makebox{$v_0$}}
    \put(235,48){\makebox{$v_1$}}
    \drawline(212,47)(212,45)
        (232,45)(232,47)(232,45)
        (247,45)(247,47)
    \dottedline{2}(212,47)(212,60)

    \put(315,48){\makebox{$v_0$}}
    \put(333,48){\makebox{$v_1$}}
    \drawline(310,47)(310,45)
        (330,45)(330,47)(330,45)
        (345,45)(345,47)
    \dottedline{2}(310,47)(310,60)

    \put(347,48){\makebox{$t'$}}
    \drawline(345,47)(345,45)
         (355,45)(355,47)
    \dottedline{2}(355,47)(355,60)

    \put(0,33){\makebox{$w_0$}}
    \put(12.6,33){\makebox{$b$}}
    \put(25,33){\makebox{$z$}}
    \put(103,33){\makebox{$w_1$}}
    \drawline(0,32)(0,30)
        (12,30)(12,32)(12,30)
        (18,30)(18,32)(18,30)
        (37,30)(37,32)(37,30)
        (180,30)(180,32)
    \dottedline{2}(0,32)(0,75)
    \dottedline{2}(180,32)(180,60)

    \put(186.5,33){\makebox{$u_0$}}
    \drawline(180,32)(180,30)(204,30)(204,32)

    \put(204.6,33){\makebox{$b$}}
    \put(217,33){\makebox{$z$}}
    \drawline(204,32)(204,30)
        (210,30)(210,32)(210,30)
        (229,30)(229,32)

    \put(297,33){\makebox{$u_1$}}
    \drawline(247,32)(247,30)(355,30)(355,32)
    \dottedline{2}(247,32)(247,45)
    \dottedline{2}(355,32)(355,45)

    \put(14,18){\makebox{$h_0$}}
    \put(206,18){\makebox{$h_0$}}
    \drawline(0,17)(0,15)
        (37,15)(37,17)(37,15)
        (192,15)(192,17)(192,15)
        (229,15)(229,17)
    \dottedline{2}(0,17)(0,30)
    \dottedline{2}(229,17)(229,30)

    \put(113,3){\makebox{$f_0$}}
  \end{picture}
  \end{center}
  Surely, $\SUF{bz}{h_0}$
  otherwise $w$ is bordered with a~suffix of~$z$. Moreover,
  $|w_0bz|\leq|h_0|$ and $|h_0|\leq|u_0 bz|$ since $bz$
  does not occur in $w_0$ and $w$ is unbordered. From
  that and $w_0h=w'_0v_0v_1$ and $u_0h=u'_0v_0v_1$
  follows now $|w'_0|\leq|u'_0|$ and
  \begin{equation}
    u'_0v_0v_1 = u_0bzv'
    \text{ and $w_0$ occurs in $u_0$.}
    \label{eq:10}
  \end{equation}

  Let now
  \[
    w = w'_0 v_0v_1 w'_i \cdots v_0v_1 w'_2 v_0v_1
        w'_1 v_0v_1 w_2
  \]
  for some word $w_2$ that does not contain $v_0v_1$, and
  \[
    u = u'_0 v_0v_1 u'_j \cdots v_0v_1 u'_2 v_0v_1
        u'_1 v_0v_1 t'
  \]
  such that $v_0v_1$ does not occur in $w'_k$,
  for all \hbox{$0\leq k\leq i$},
  or $v'_\ell$, for all $0\leq\ell\leq j$.
  Note, that these factorizations of $w$ and $u$
  are unique, and, moreover, $w_2\neq\eps$. (Indeed,
  if $w_2=\eps$ then $\SUF{v_0v_1}{w}$
  and $\SUF{az}{v_0v_1}$, and $az$ would occur
  in $u$; a~contradiction.)

  We claim that either $i=j$ and $w'_k=u'_k$,
  for all $1\leq k\leq i$ or $|u| < |w| - 1$.

  Assume $k=1$. We show that $w'_1=u'_1$. Consider
  \[
    f_1 = v_1  w'_1 v_0v_1 w_2 u'_0 v_0v_1 u'_j \cdots v_0v_1
          u'_1 v_0 \ .
  \]
  If $f_1$ is unbordered, then $|u|<|w|-1$ since
  $|f_1|\leq|w|$ and
  \[
    |u| = |f_1| - |v_1  w'_1 v_0v_1 w_2| + |v_1 t'|
  \]
  and $|t'|\leq|z_0|\leq|z|<|bz|\leq|v_0v_1|$ and $w_2\neq\eps$.
  Assume then that $f_1$ is bordered, and let
  $h_1$ be its shortest border. Clearly, $h_1=v_1 g_1 v_0$
  for some $g_1$ (possibly $g_1=\eps$) since $v_0$ and $v_1$
  do not overlap.
  We show that $\PRE{h_1}{v_1w'_1v_0}$. Indeed, otherwise either
  \begin{enumerate}
    \item $az$ occurs in $u$, in case $\PRE{v_1w'_1v_0v_1w_2}{h_1}$,
      a contradiction to our assumption on~$az$, or
    \item $v_0$ and $v_1$ overlap, in case $|v_0|\leq|z|$ and
      \[
        |v_1w'_1v_0v_1w_2|-|az|+|v_0|<|h_1|<|v_1w'_1v_0v_1w_2|
      \]
      and then $v_0$ occurs in $z$, contradicting Lemma~\ref{lem:6}, or
    \item $|u|<|w|-1$, in case $\SUF{v_0w_3}{w_2}$
      and $|az|\leq|v_0w_3|$, then $v_0w_3u'v_0v_1$ is unbordered
      and the result follows from $|t'|<|v_0w_3|-1$,
      since \hbox{$|az|\neq|v_0w_3|$} for $v_0$ does not begin with $a$.
  \end{enumerate}
  Moreover, $\SUF{h_1}{v_1u'_1v_0}$ since $v_0v_1$ does
  not occur in $v_1w'_1v_0$. So, let
  \begin{equation}\label{eq:2}
    w'_1 v_0 = g_1 v_0 w''_1
    \qquad\text{and}\qquad
    v_1 u'_1 = u''_1 v_1 g_1 \ .
  \end{equation}
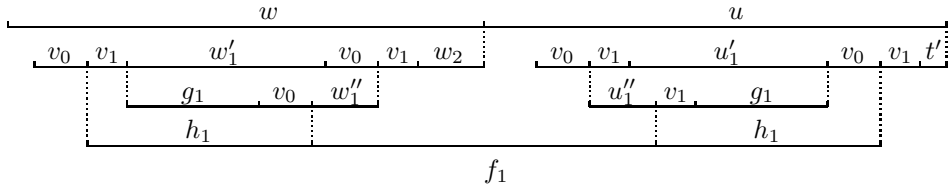
  \begin{center}
  \begin{picture}(355,70)(0,0)
    \put(95,63){\makebox{$w$}}
    \put(272.5,63){\makebox{$u$}}
    \drawline(0,62)(0,60)(180,60)(180,62)
    \drawline(180,62)(180,60)(355,60)(355,62)

    \put(15,48){\makebox{$v_0$}}
    \put(33,48){\makebox{$v_1$}}
    \put(76,48){\makebox{$w'_1$}}
    \put(125,48){\makebox{$v_0$}}
    \put(143,48){\makebox{$v_1$}}
    \put(161,48){\makebox{$w_2$}}
    \drawline(10,47)(10,45)
        (30,45)(30,47)(30,45)
        (45,45)(45,47)(45,45)
        (120,45)(120,47)(120,45)
        (140,45)(140,47)(140,45)
        (155,45)(155,47)(155,45)
        (180,45)(180,47)
    \dottedline{2}(180,47)(180,60)

    \put(205,48){\makebox{$v_0$}}
    \put(223,48){\makebox{$v_1$}}
    \put(267,48){\makebox{$u'_1$}}
    \put(315,48){\makebox{$v_0$}}
    \put(333,48){\makebox{$v_1$}}
    \put(347,48){\makebox{$t'$}}
    \drawline(200,47)(200,45)
        (220,45)(220,47)(220,45)
        (235,45)(235,47)(235,45)
        (310,45)(310,47)(310,45)
        (330,45)(330,47)(330,45)
        (345,45)(345,47)(345,45)
         (355,45)(355,47)
    \dottedline{2}(355,47)(355,60)

    \put(65.5,33){\makebox{$g_1$}}
    \put(100,33){\makebox{$v_0$}}
    \put(122,33){\makebox{$w''_1$}}
    \drawline(45,32)(45,30)
        (95,30)(95,32)(95,30)
        (115,30)(115,32)(115,30)
        (140,30)(140,32)
    \dottedline{2}(45,32)(45,45)
    \dottedline{2}(140,32)(140,45)

    \put(227,33){\makebox{$u''_1$}}
    \put(248,33){\makebox{$v_1$}}
    \put(280.5,33){\makebox{$g_1$}}
    \drawline(220,32)(220,30)
        (245,30)(245,32)(245,30)
        (260,30)(260,32)(260,30)
        (310,30)(310,32)
    \dottedline{2}(220,32)(220,45)
    \dottedline{2}(310,32)(310,45)

    \put(67,18){\makebox{$h_1$}}
    \put(282,18){\makebox{$h_1$}}
    \drawline(30,17)(30,15)
        (115,15)(115,17)(115,15)
        (245,15)(245,17)(245,15)
        (330,15)(330,17)
    \dottedline{2}(30,17)(30,45)
    \dottedline{2}(115,17)(115,30)
    \dottedline{2}(245,17)(245,30)
    \dottedline{2}(330,17)(330,45)

    \put(180,3){\makebox{$f_1$}}
  \end{picture}
  \end{center}

  Consider,
  \[
    f_2 = v_0 w''_1 v_1 w_2 u'_0 v_0v_1 u'_j \cdots v_0v_1
          u'_1 v_0v_1 \ .
  \]
  If $f_2$ is unbordered, then $|u|<|w|-1$ since
  $|f_2|\leq|w|$ and
  \[
    |u| = |f_2| - |v_0 w''_1 v_1 w_2| + |t'|
  \]
  and $|t'|\leq|z_0|\leq|z|<|bz|\leq|v_0v_1|$ and $w_2\neq\eps$.
  Assume then that $f_2$ is bordered, and let
  $h_2$ be its shortest border. Since $v_0$ and $v_1$ do not
  overlap, $\SUF{v_0v_1}{h_2}$.
  Also \hbox{$\PRE{h_2}{v_0 w''_1 v_1}$} since $v_0v_1$ does
  not occur in $w_2$ (and $v_0$ and $v_1$ do not overlap)
  and $az$ does not occur in $h_2$ (and so $h_2$
  does not stretch beyond $w$). We have
  $\PRE{v_0 w''_1 v_1}{h_2}$ since $v_0v_1$ does not occur
  in~$v_0w''_1v_1$ unless $w''_1=\eps$.
  Hence, we have $h_2=v_0 w''_1 v_1$ and
  \begin{equation}
    w'_1 v_0v_1 = g_1 h_2
    \qquad\text{and}\qquad
    \SUF{h_2}{u'_1 v_0v_1} \ . \label{eq:3}
  \end{equation}
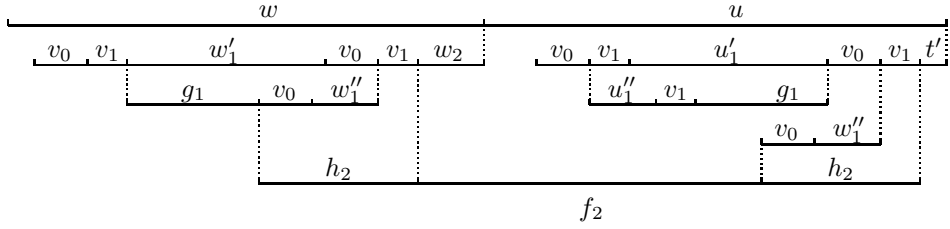
  \begin{center}
  \begin{picture}(355,85)(0,0)
    \put(95,78){\makebox{$w$}}
    \put(272.5,78){\makebox{$u$}}
    \drawline(0,77)(0,75)(180,75)(180,77)
    \drawline(180,77)(180,75)(355,75)(355,77)

    \put(15,63){\makebox{$v_0$}}
    \put(33,63){\makebox{$v_1$}}
    \put(76,63){\makebox{$w'_1$}}
    \put(125,63){\makebox{$v_0$}}
    \put(143,63){\makebox{$v_1$}}
    \put(161,63){\makebox{$w_2$}}
    \drawline(10,62)(10,60)
        (30,60)(30,62)(30,60)
        (45,60)(45,62)(45,60)
        (120,60)(120,62)(120,60)
        (140,60)(140,62)(140,60)
        (155,60)(155,62)(155,60)
        (180,60)(180,62)
    \dottedline{2}(180,62)(180,75)

    \put(205,63){\makebox{$v_0$}}
    \put(223,63){\makebox{$v_1$}}
    \put(267,63){\makebox{$u'_1$}}
    \put(315,63){\makebox{$v_0$}}
    \put(333,63){\makebox{$v_1$}}
    \put(347,63){\makebox{$t'$}}
    \drawline(200,62)(200,60)
        (220,60)(220,62)(220,60)
        (235,60)(235,62)(235,60)
        (310,60)(310,62)(310,60)
        (330,60)(330,62)(330,60)
        (345,60)(345,62)(345,60)
        (355,60)(355,62)
    \dottedline{2}(355,62)(355,75)

    \put(65.5,48){\makebox{$g_1$}}
    \put(100,48){\makebox{$v_0$}}
    \put(122,48){\makebox{$w''_1$}}
    \drawline(45,47)(45,45)
        (95,45)(95,47)(95,45)
        (115,45)(115,47)(115,45)
        (140,45)(140,47)
    \dottedline{2}(45,47)(45,60)
    \dottedline{2}(140,47)(140,60)

    \put(227,48){\makebox{$u''_1$}}
    \put(248,48){\makebox{$v_1$}}
    \put(290.5,48){\makebox{$g_1$}}
    \drawline(220,47)(220,45)
        (245,45)(245,47)(245,45)
        (260,45)(260,47)(260,45)
        (310,45)(310,47)
    \dottedline{2}(220,47)(220,60)
    \dottedline{2}(310,47)(310,60)

    \put(290,33){\makebox{$v_0$}}
    \put(312,33){\makebox{$w''_1$}}
    \drawline(285,32)(285,30)
        (305,30)(305,32)(305,30)
        (330,30)(330,32)
    \dottedline{2}(330,32)(330,60)

    \put(120,18){\makebox{$h_2$}}
    \put(310,18){\makebox{$h_2$}}
    \drawline(95,17)(95,15)
        (155,15)(155,17)(155,15)
        (285,15)(285,17)(285,15)
        (345,15)(345,17)
    \dottedline{2}(95,17)(95,45)
    \dottedline{2}(155,17)(155,60)
    \dottedline{2}(285,17)(285,30)
    \dottedline{2}(345,17)(345,60)

    \put(216,3){\makebox{$f_2$}}
  \end{picture}
  \end{center}

  Consider,
  \[
    f_3 = v_0v_1 w'_1 v_0v_1 w_2 u'_0 v_0v_1 u'_j \cdots v_0v_1
          u'_2 v_0 u''_1 v_1 \ .
  \]
  If $f_3$ is unbordered, then $|u|<|w|-1$ since
  $|f_3|\leq|w|$ and
  \[
    |u| = |f_3| - |v_0v_1 w'_1 v_0v_1 w_2| + |g_1 v_0v_1 t'|
  \]
  and $|t'|\leq|z_0|\leq|z|<|bz|\leq|v_0v_1|$ and $|g_1|\leq|w'_1|$
  and $w_2\neq\eps$.
  Assume, that $f_3$ is bordered. Then
  $f_3$ has a shortest border $h_3$ such that $\PRE{v_0v_1}{h_3}$.
  We have $h_3=v_0u''_1v_1$ by the arguments from the previous
  paragraph. Moreover,
  \begin{equation}
    v_0v_1 u'_1 = h_3 g_1
    \qquad\text{and}\qquad
    \PRE{v_0v_1w'_1}{h_3} \ . \label{eq:4}
  \end{equation}
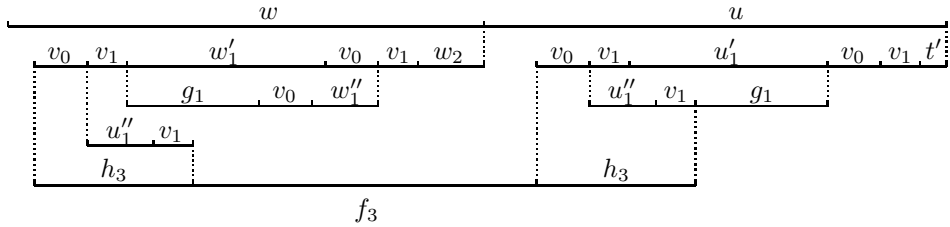
  \begin{center}
  \begin{picture}(355,85)(0,0)
    \put(95,78){\makebox{$w$}}
    \put(272.5,78){\makebox{$u$}}
    \drawline(0,77)(0,75)(180,75)(180,77)
    \drawline(180,77)(180,75)(355,75)(355,77)

    \put(15,63){\makebox{$v_0$}}
    \put(33,63){\makebox{$v_1$}}
    \put(76,63){\makebox{$w'_1$}}
    \put(125,63){\makebox{$v_0$}}
    \put(143,63){\makebox{$v_1$}}
    \put(161,63){\makebox{$w_2$}}
    \drawline(10,62)(10,60)
        (30,60)(30,62)(30,60)
        (45,60)(45,62)(45,60)
        (120,60)(120,62)(120,60)
        (140,60)(140,62)(140,60)
        (155,60)(155,62)(155,60)
        (180,60)(180,62)
    \dottedline{2}(180,62)(180,75)

    \put(205,63){\makebox{$v_0$}}
    \put(223,63){\makebox{$v_1$}}
    \put(267,63){\makebox{$u'_1$}}
    \put(315,63){\makebox{$v_0$}}
    \put(333,63){\makebox{$v_1$}}
    \put(347,63){\makebox{$t'$}}
    \drawline(200,62)(200,60)
        (220,60)(220,62)(220,60)
        (235,60)(235,62)(235,60)
        (310,60)(310,62)(310,60)
        (330,60)(330,62)(330,60)
        (345,60)(345,62)(345,60)
        (355,60)(355,62)
    \dottedline{2}(355,62)(355,75)

    \put(65.5,48){\makebox{$g_1$}}
    \put(100,48){\makebox{$v_0$}}
    \put(122,48){\makebox{$w''_1$}}
    \drawline(45,47)(45,45)
        (95,45)(95,47)(95,45)
        (115,45)(115,47)(115,45)
        (140,45)(140,47)
    \dottedline{2}(45,47)(45,60)
    \dottedline{2}(140,47)(140,60)

    \put(227,48){\makebox{$u''_1$}}
    \put(248,48){\makebox{$v_1$}}
    \put(280.5,48){\makebox{$g_1$}}
    \drawline(220,47)(220,45)
        (245,45)(245,47)(245,45)
        (260,45)(260,47)(260,45)
        (310,45)(310,47)
    \dottedline{2}(220,47)(220,60)
    \dottedline{2}(310,47)(310,60)

    \put(57,33){\makebox{$v_1$}}
    \put(37,33){\makebox{$u''_1$}}
    \drawline(30,32)(30,30)
        (55,30)(55,32)(55,30)
        (70,30)(70,32)
    \dottedline{2}(30,32)(30,60)

    \put(35,18){\makebox{$h_3$}}
    \put(225,18){\makebox{$h_3$}}
    \drawline(10,17)(10,15)
        (70,15)(70,17)(70,15)
        (200,15)(200,17)(200,15)
        (260,15)(260,17)
    \dottedline{2}(10,17)(10,60)
    \dottedline{2}(70,17)(70,30)
    \dottedline{2}(200,17)(200,60)
    \dottedline{2}(260,17)(260,45)

    \put(131,3){\makebox{$f_3$}}
  \end{picture}
  \end{center}

  Observe, that~(\ref{eq:3}) and~(\ref{eq:4}) imply
  that the number of occurrences of $v_1$ and $v_0$, respectively,
  is the same in $w'_1$ and $u'_1$ since $v_0$ and $v_1$ do not
  overlap. Now, let
  \[
    h_1 = v_1 g_1 v_0 = h''_1 v_1 h'_1 v_0 = v_1 h'_0 v_0 h''_0
  \]
  where $v_1$ and $v_0$ occur only once in $v_1 h'_1$ and $h'_0 v_0$,
  respectively.
  
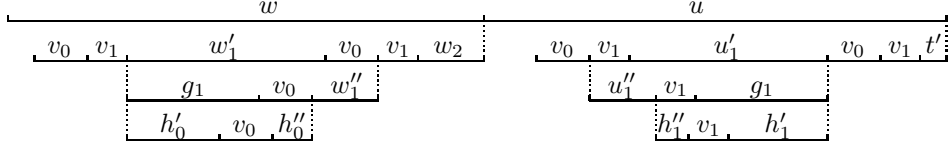
\begin{center}
  \begin{picture}(355,55)(0,0)
    \put(95,48){\makebox{$w$}}
    \put(257.5,48){\makebox{$u$}}
    \drawline(0,47)(0,45)(180,45)(180,47)
    \drawline(180,47)(180,45)(355,45)(355,47)

    \put(15,33){\makebox{$v_0$}}
    \put(33,33){\makebox{$v_1$}}
    \put(76,33){\makebox{$w'_1$}}
    \put(125,33){\makebox{$v_0$}}
    \put(143,33){\makebox{$v_1$}}
    \put(161,33){\makebox{$w_2$}}
    \drawline(10,32)(10,30)
        (30,30)(30,32)(30,30)
        (45,30)(45,32)(45,30)
        (120,30)(120,32)(120,30)
        (140,30)(140,32)(140,30)
        (155,30)(155,32)(155,30)
        (180,30)(180,32)
    \dottedline{2}(180,32)(180,45)

    \put(205,33){\makebox{$v_0$}}
    \put(223,33){\makebox{$v_1$}}
    \put(267,33){\makebox{$u'_1$}}
    \put(315,33){\makebox{$v_0$}}
    \put(333,33){\makebox{$v_1$}}
    \put(347,33){\makebox{$t'$}}
    \drawline(200,32)(200,30)
        (220,30)(220,32)(220,30)
        (235,30)(235,32)(235,30)
        (310,30)(310,32)(310,30)
        (330,30)(330,32)(330,30)
        (345,30)(345,32)(345,30)
        (355,30)(355,32)
    \dottedline{2}(355,32)(355,45)

    \put(65.5,18){\makebox{$g_1$}}
    \put(100,18){\makebox{$v_0$}}
    \put(122,18){\makebox{$w''_1$}}
    \drawline(45,17)(45,15)
        (95,15)(95,17)(95,15)
        (115,15)(115,17)(115,15)
        (140,15)(140,17)
    \dottedline{2}(45,17)(45,30)
    \dottedline{2}(140,17)(140,30)

    \put(227,18){\makebox{$u''_1$}}
    \put(248,18){\makebox{$v_1$}}
    \put(280.5,18){\makebox{$g_1$}}
    \drawline(220,17)(220,15)
        (245,15)(245,17)(245,15)
        (260,15)(260,17)(260,15)
        (310,15)(310,17)
    \dottedline{2}(220,17)(220,30)
    \dottedline{2}(310,17)(310,30)

    \put(57.5,3){\makebox{$h'_0$}}
    \put(85,3){\makebox{$v_0$}}
    \put(102,3){\makebox{$h''_0$}}
    \drawline(45,2)(45,0)
        (80,0)(80,2)(80,0)
        (100,0)(100,2)(100,0)
        (115,0)(115,2)
    \dottedline{2}(45,2)(45,15)
    \dottedline{2}(115,2)(115,15)

    \put(245.5,3){\makebox{$h''_1$}}
    \put(260.5,3){\makebox{$v_1$}}
    \put(286.25,3){\makebox{$h'_1$}}
    \drawline(245,2)(245,0)
        (257.5,0)(257.5,2)(257.5,0)
        (272.5,0)(272.5,2)(272.5,0)
        (310,0)(310,2)
    \dottedline{2}(245,2)(245,15)
    \dottedline{2}(310,2)(310,15)
  \end{picture}
  \end{center}

  Now, let
  \[
    f'_2 = v_0 h''_0 w''_1 v_1 w_2 u'_0 v_0v_1 u'_j \cdots v_0v_1
           u'_1 v_0v_1
  \]
  and
  \[
    f'_3 = v_0v_1 w'_1 v_0v_1 w_2 u'_0 v_0v_1 u'_j \cdots v_0v_1
           u'_2 v_0 u''_1 h''_1 v_1
  \]
  with the respective shortest borders $h'_2$ and $h'_3$ (which
  are both not empty, if \hbox{$|u|\geq|w|-1$}; as in the case
  of $f_2$ and $f_3$) and $\SUF{v_0v_1}{h'_2}$
  and $\PRE{v_0v_1}{h'_3}$.

  We have \hbox{$\PRE{h'_2}{v_0 h''_0 w''_1 v_1}$} since $v_0v_1$
  does not occur in $w_2$ and $az$ does not occur in $h'_2$
  (and so $h'_2$ does not stretch beyond $w$). We have
  $\PRE{v_0 h''_0 w''_1 v_1}{h'_2}$ since $v_0v_1$ does not occur
  in~$w'_1$. Hence, we have $h'_2=v_0 h''_0 w''_1 v_1$ and
  \[
    w'_1 v_0v_1 = h'_0 v_0 h''_2 w''_1 v_1 = h'_0 h'_2
    \qquad\text{and}\qquad
    \SUF{h'_2}{u'_1 v_0v_1} \ .
  \]
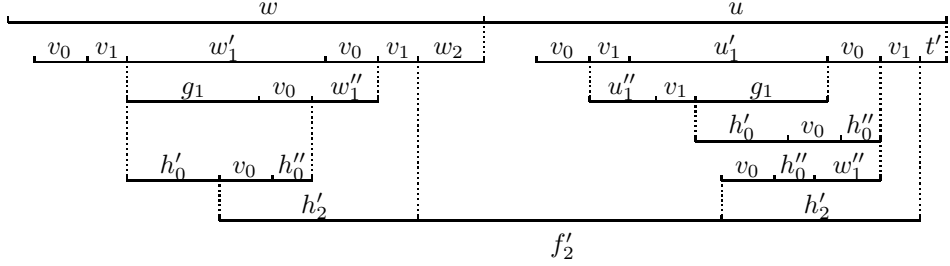
  \begin{center}
  \begin{picture}(355,100)(0,0)
    \put(95,93){\makebox{$w$}}
    \put(272.5,93){\makebox{$u$}}
    \drawline(0,92)(0,90)(180,90)(180,92)
    \drawline(180,92)(180,90)(355,90)(355,92)

    \put(15,78){\makebox{$v_0$}}
    \put(33,78){\makebox{$v_1$}}
    \put(76,78){\makebox{$w'_1$}}
    \put(125,78){\makebox{$v_0$}}
    \put(143,78){\makebox{$v_1$}}
    \put(161,78){\makebox{$w_2$}}
    \drawline(10,77)(10,75)
        (30,75)(30,77)(30,75)
        (45,75)(45,77)(45,75)
        (120,75)(120,77)(120,75)
        (140,75)(140,77)(140,75)
        (155,75)(155,77)(155,75)
        (180,75)(180,77)
    \dottedline{2}(180,77)(180,90)

    \put(205,78){\makebox{$v_0$}}
    \put(223,78){\makebox{$v_1$}}
    \put(267,78){\makebox{$u'_1$}}
    \put(315,78){\makebox{$v_0$}}
    \put(333,78){\makebox{$v_1$}}
    \put(347,78){\makebox{$t'$}}
    \drawline(200,77)(200,75)
        (220,75)(220,77)(220,75)
        (235,75)(235,77)(235,75)
        (310,75)(310,77)(310,75)
        (330,75)(330,77)(330,75)
        (345,75)(345,77)(345,75)
        (355,75)(355,77)
    \dottedline{2}(355,77)(355,90)

    \put(65.5,63){\makebox{$g_1$}}
    \put(100,63){\makebox{$v_0$}}
    \put(122,63){\makebox{$w''_1$}}
    \drawline(45,62)(45,60)
        (95,60)(95,62)(95,60)
        (115,60)(115,62)(115,60)
        (140,60)(140,62)
    \dottedline{2}(45,62)(45,75)
    \dottedline{2}(140,62)(140,75)

    \put(227,63){\makebox{$u''_1$}}
    \put(248,63){\makebox{$v_1$}}
    \put(280.5,63){\makebox{$g_1$}}
    \drawline(220,62)(220,60)
        (245,60)(245,62)(245,60)
        (260,60)(260,62)(260,60)
        (310,60)(310,62)
    \dottedline{2}(220,62)(220,75)
    \dottedline{2}(310,62)(310,75)

    \put(57.5,33){\makebox{$h'_0$}}
    \put(85,33){\makebox{$v_0$}}
    \put(102,33){\makebox{$h''_0$}}
    \drawline(45,32)(45,30)
        (80,30)(80,32)(80,30)
        (100,30)(100,32)(100,30)
        (115,30)(115,32)
    \dottedline{2}(45,32)(45,60)
    \dottedline{2}(115,32)(115,60)

    \put(272.5,48){\makebox{$h'_0$}}
    \put(300,48){\makebox{$v_0$}}
    \put(317,48){\makebox{$h''_0$}}
    \drawline(260,47)(260,45)
        (295,45)(295,47)(295,45)
        (315,45)(315,47)(315,45)
        (330,45)(330,47)
    \dottedline{2}(260,47)(260,60)
    \dottedline{2}(330,47)(330,75)

    \put(275,33){\makebox{$v_0$}}
    \put(292,33){\makebox{$h''_0$}}
    \put(312,33){\makebox{$w''_1$}}
    \drawline(270,32)(270,30)
        (290,30)(290,32)(290,30)
        (305,30)(305,32)(305,30)
        (330,30)(330,32)
    \dottedline{2}(330,32)(330,45)

    \put(111,18){\makebox{$h'_2$}}
    \put(301,18){\makebox{$h'_2$}}
    \drawline(80,17)(80,15)
        (155,15)(155,17)(155,15)
        (270,15)(270,17)(270,15)
        (345,15)(345,17)
    \dottedline{2}(80,17)(80,30)
    \dottedline{2}(155,17)(155,75)
    \dottedline{2}(270,17)(270,30)
    \dottedline{2}(345,17)(345,75)

    \put(205,3){\makebox{$f'_2$}}
  \end{picture}
  \end{center}

  We have $h'_3=v_0u''_1h''_1v_1$ by the arguments from the previous
  paragraph. Moreover,
  \[
    v_0v_1 u'_1 = v_0 u''_1 h''_1 v_1 h'_1 = h'_3 h'_1
    \qquad\text{and}\qquad
    \PRE{v_0v_1w'_1}{h'_3} \ .
  \]
  \begin{center}
  \begin{picture}(355,100)(0,0)
    \put(95,93){\makebox{$w$}}
    \put(272.5,93){\makebox{$u$}}
    \drawline(0,92)(0,90)(180,90)(180,92)
    \drawline(180,92)(180,90)(355,90)(355,92)

    \put(15,78){\makebox{$v_0$}}
    \put(33,78){\makebox{$v_1$}}
    \put(76,78){\makebox{$w'_1$}}
    \put(125,78){\makebox{$v_0$}}
    \put(143,78){\makebox{$v_1$}}
    \put(161,78){\makebox{$w_2$}}
    \drawline(10,77)(10,75)
        (30,75)(30,77)(30,75)
        (45,75)(45,77)(45,75)
        (120,75)(120,77)(120,75)
        (140,75)(140,77)(140,75)
        (155,75)(155,77)(155,75)
        (180,75)(180,77)
    \dottedline{2}(180,77)(180,90)

    \put(205,78){\makebox{$v_0$}}
    \put(223,78){\makebox{$v_1$}}
    \put(267,78){\makebox{$u'_1$}}
    \put(315,78){\makebox{$v_0$}}
    \put(333,78){\makebox{$v_1$}}
    \put(347,78){\makebox{$t'$}}
    \drawline(200,77)(200,75)
        (220,75)(220,77)(220,75)
        (235,75)(235,77)(235,75)
        (310,75)(310,77)(310,75)
        (330,75)(330,77)(330,75)
        (345,75)(345,77)(345,75)
        (355,75)(355,77)
    \dottedline{2}(355,77)(355,90)

    \put(65.5,63){\makebox{$g_1$}}
    \put(100,63){\makebox{$v_0$}}
    \put(122,63){\makebox{$w''_1$}}
    \drawline(45,62)(45,60)
        (95,60)(95,62)(95,60)
        (115,60)(115,62)(115,60)
        (140,60)(140,62)
    \dottedline{2}(45,62)(45,75)
    \dottedline{2}(140,62)(140,75)

    \put(227,63){\makebox{$u''_1$}}
    \put(248,63){\makebox{$v_1$}}
    \put(280.5,63){\makebox{$g_1$}}
    \drawline(220,62)(220,60)
        (245,60)(245,62)(245,60)
        (260,60)(260,62)(260,60)
        (310,60)(310,62)
    \dottedline{2}(220,62)(220,75)
    \dottedline{2}(310,62)(310,75)

    \put(30.5,48){\makebox{$h''_1$}}
    \put(45.5,48){\makebox{$v_1$}}
    \put(71.25,48){\makebox{$h'_1$}}
    \drawline(30,47)(30,45)
        (42.5,45)(42.5,47)(42.5,45)
        (57.5,45)(57.5,47)(57.5,45)
        (95,45)(95,47)
    \dottedline{2}(30,47)(30,75)
    \dottedline{2}(95,47)(95,60)

    \put(37,33){\makebox{$u''_1$}}
    \put(55.5,33){\makebox{$h''_1$}}
    \put(70.5,33){\makebox{$v_1$}}
    \drawline(30,32)(30,30)
        (55,30)(55,32)(55,30)
        (67.5,30)(67.5,32)(67.5,30)
        (82.5,30)(82.5,32)
    \dottedline{2}(30,32)(30,45)

    \put(245.5,33){\makebox{$h''_1$}}
    \put(260.5,33){\makebox{$v_1$}}
    \put(286.25,33){\makebox{$h'_1$}}
    \drawline(245,32)(245,30)
        (257.5,30)(257.5,32)(257.5,30)
        (272.5,30)(272.5,32)(272.5,30)
        (310,30)(310,32)
    \dottedline{2}(245,32)(245,60)
    \dottedline{2}(310,32)(310,60)

    \put(40,18){\makebox{$h'_3$}}
    \put(230,18){\makebox{$h'_3$}}
    \drawline(10,17)(10,15)
        (82.5,15)(82.5,17)(82.5,15)
        (200,15)(200,17)(200,15)
        (272.5,15)(272.5,17)
    \dottedline{2}(10,17)(10,75)
    \dottedline{2}(82.5,17)(82.5,30)
    \dottedline{2}(200,17)(200,75)
    \dottedline{2}(272.5,17)(272.5,30)

    \put(130,3){\makebox{$f'_3$}}
  \end{picture}
  \end{center}
  It is now straightforward to see that
  \[
    w''_1 = u''_1 = \eps
  \]
  for otherwise $v_1$ and $v_0$ occur more than once in $v_1h'_1$
  and $h'_0v_0$, respectively. From~\eqref{eq:2} follows now
  \[
    w'_1 = g_1 = u'_1 \ .
  \]

  Assume $1<k\leq\min\{i, j\}$ and $w'_\ell=u'_\ell$,
  for all $1\leq\ell<k$. Let us denote both $w'_\ell$
  and $u'_\ell$ by $v'_\ell$, for all $1\leq\ell<k$.

  We show that $w'_k=u'_k$. Consider
  \[
    f_4 = v_1 w'_k v_0v_1 v'_{k-1} v_0v_1 \cdots v'_1 v_0v_1 w_2
          u'_0 v_0v_1 u'_j \cdots v_0v_1 u'_k v_0 \ .
  \]
  If $f_4$ is unbordered, then $|u|<|w|-1$ since
  $|f_4|\leq|w|$ and
  \[
    |u| = |f_4| - |v_1 w'_kv_0v_1v'_{k-1}v_0v_1\cdots v'_1v_0v_1w_2|
                + |v_1 v'_{k-1}v_0v_1\cdots v'_1 v_0v_1 t'|
  \]
  and $|t'|\leq|z_0|\leq|z|<|bz|\leq|v_0v_1|$ and $w_2\neq\eps$.
  Assume, $f_4$ is bordered. Then
  $f_4$ has a shortest border $h_4$ such that $|v_0v_1|\leq|h_4|$.
  Let $h_4=v_1 g_4 v_0$.

  If $|v_1w'_kv_0|<|h_4|$ then there exists an $\ell<k$ such
  that
  \[
    h_4 = v_1 w'_k v_0v_1 v'_{k-1} v_0v_1 \cdots
          v'_{\ell+1} v_0v_1 v''_\ell v_0 
  \]
  where $\PRE{v''_\ell}{v'_\ell}$. That implies
  \[
    u'_k = v''_\ell
  \]
  since $v_0v_1$ does neither occur in $v''_\ell$ nor in $u'_k$.
  Now, consider
  \[
    f_5 = v_1 w'_k v_0v_1 v'_{k-1} v_0v_1 \cdots v'_1 v_0v_1 w_2
          u'_0 v_0v_1 u'_j \cdots v_0v_1 u'_k v_0v_1 v'_{k-1} v_0v_1
          \cdots v''_\ell v_0 \ .
  \]
  If $f_5$ is unbordered, then $|u|<|w|-1$ since $|f_4|<|f_5|$,
  see above.
  Assume, $f_5$ is bordered. Then
  $f_5$ has a shortest border $h_5$ such that
  \[
    |h_4| < |h_5|
  \]
  for otherwise $h_4$ is not the shortest border of $f_4$,
  since either $\PRE{h_4}{h_5}$ or $\PRE{h_5}{h_4}$,
  and the latter implies that $h_4$ is bordered,
  and hence, not minimal. But now, we have a $\ell'<\ell$
  such that
  \[
    h_5 = v_1 w'_k v_0v_1 v'_{k-1} v_0v_1 \cdots
          v'_{\ell'+1} v_0v_1 v''_{\ell'} v_0 
  \]
  where $\PRE{v''_{\ell'}}{v'_{\ell'}}$.
  We have $|f_4|<|f_5|<|f_6|$ where
  \[
    f_6 = v_1 w'_k v_0v_1 v'_{k-1} v_0v_1 \cdots v'_1 v_0v_1 w_2
          u'_0 v_0v_1 u'_j \cdots v_0v_1 u'_k v_0v_1 v'_{k-1} v_0v_1
          \cdots v''_{\ell'} v_0 \ ,
  \]
  which is either unbordered and $|u|<|w|-1$ since $|f_4|<|f_5|$,
  or it is bordered with a shortest border $h_6$,
  and we have $|h_4|<|h_5|<|h_6|$ and a factor $f_7$,
  such that $|f_4|<|f_5|<|f_6|<|f_7|$, and so on, until eventually
  an unbordered factor is reached proving that $|u|<|w|-1$.

  Assume then that $\PRE{h_4}{v_1w'_kv_0}$. We also have that
  $\SUF{h_4}{v_1u'_kv_0}$ since $v_0v_1$ does not occur
  in $w'_k$. So, let $w'_kv_0=g_4v_0w''_k$
  and $v_1u'_k=u''_kv_1g_4$.

  Consider,
  \[
    f_8 = v_0 w''_k v_1 v'_{k-1} v_0v_1 \cdots
             v'_1 v_0v_1 w_2 u'_0 v_0v_1 u'_j v_0v_1 \cdots
             u'_k v_0v_1 \ .
  \]
  If $f_8$ is unbordered, then $|u|<|w|-1$ since
  $|f_8|\leq|w|$ and
  \[
    |u| = |f_8| - |v_0 w''_k v_1v'_{k-1}v_0v_1\cdots v'_1v_0v_1w_2|
                + |v'_{k-1}v_0v_1\cdots v'_1 v_0v_1 t'|
  \]
  and $|t'|\leq|z_0|\leq|z|<|bz|\leq|v_0v_1|$ and $w_2\neq\eps$.
  Assume, $f_8$ is bordered. Then
  $f_8$ has a shortest border $h_8$ such that
  $\SUF{v_0v_1}{h_8}$.

  If $|h_8|>|v_0 w''_k v_1|$ then the same argument as
  in the case $|v_1w'_kv_0|<|h_4|$ above shows that
  $|u|<|w|-1$.
  If $|h_8|<|v_0 w''_k v_1|$ then
  $v_0v_1$ occurs in $w'_k$; a~contradiction.
  Hence, we have $h_8=v_0 w''_k v_1$ and
  \begin{equation}
    w'_k v_0v_1 = g_1 h_8 
    \qquad\text{and}\qquad
    \SUF{h_8}{u'_kv_0v_1} \ . \label{eq:5}
  \end{equation}

  Consider,
  \[
    f_9 = v_0v_1 w'_k v_0v_1 v'_{k-1} v_0v_1 \cdots
             v'_1 v_0v_1 w_2 u'_0 v_0v_1 u'_j v_0v_1 \cdots
             u'_{k+1} v_0 u''_k v_1 \ .
  \]
  If $f_9$ is unbordered, then $|u|<|w|-1$ since
  $|f_9|\leq|w|$ and
  \[
    |u| = |f_9| - |v_0v_1w'_kv_0v_1v'_{k-1}v_0v_1\cdots v'_1v_0v_1w_2|
                + |g_4v_0v_1v'_{k-1}v_0v_1\cdots v'_1 v_0v_1 t'|
  \]
  and $|t'|\leq|z_0|\leq|z|<|bz|\leq|v_0v_1|$ and $|g_4|\leq|w'_k|$
  and $w_2\neq\eps$.
  Assume, $f_9$ is bordered. Then
  $f_9$ has a shortest border $h_9$ such
  that $\PRE{v_0v_1}{h_9}$. We have $h_9=v_0u''_kv_1$
  by the arguments from the previous paragraph. Moreover,
  \begin{equation}
    v_0v_1 u'_k = h_9 g_1 
    \qquad\text{and}\qquad
    \PRE{h_9}{v_0v_1w'_k} \ . \label{eq:6}
  \end{equation}

  Observe, that~(\ref{eq:5}) and~(\ref{eq:6}) imply
  that the number of occurrences of $v_1$ and $v_0$, respectively,
  is the same in $w'_k$ and $u'_k$ since $v_0$ and $v_1$ do not
  overlap. Now, let
  \[
    h_4 = v_1 g_4 v_0 = h''_1 v_1 h'_1 v_0 = v_1 h'_0 v_0 h''_0
  \]
  where $v_1$ and $v_0$ occur only once in $v_1 h'_1$ and $h'_0 v_0$,
  respectively.

  Now, let
  \[
    f'_8 = v_0 h''_0 w''_k v_1 v'_{k-1} \cdots
              v_0v_1 v'_1 v_0v_1 w_2 . u'_0 v_0v_1 u'_j \cdots
              v_0v_1 u'_k v_0v_1
  \]
  and
  \[
    f'_9 = v_0v_1 w'_k v_0v_1 v'_{k-1} \cdots
              v_0v_1 v'_1 v_0v_1 w_2 . u'_0 v_0v_1 u'_j \cdots
              v_0v_1 u'_{k+1} v_0 u''_1 h''_1 v_1
  \]
  with the respective shortest borders $h'_8$ and $h'_9$ (which
  are both not empty, if \hbox{$|u|\geq|w|-1$}; as in the case
  of $f_8$ and $f_9$).
  Analogously to the cases of $f_8$ and~$f_9$, we have
  \[
    w'_k v_0 v_1 = h'_0 h'_8
    \qquad\text{and}\qquad
    v_0 v_1 u'_k = h'_9 h'_1 \ .
  \]
  It is now straightforward to see that
  \[
    h'_8 = h'_9 = v_0 v_1
  \]
  and
  \[
    h_4 = v_0 w'_k v_1 = v_0 u'_k v_1
  \]
  and hence, $w'_k = u'_k$. In this case,
  we denote both~$w'_k$ and~$u'_k$ by~$v'_k$.

  Now, we have
  \begin{align*}
    \bar v &= v_0v_1 w'_\iota \cdots v_0v_1 w'_2 v_0v_1 w'_1 \\
           &= v_0v_1 u'_\iota \cdots v_0v_1 u'_2 v_0v_1 u'_1
  \end{align*}
  where $\iota=\min\{i, j\}$.

  If $i<j$ then
  \begin{equation}\label{eq:9}
    |w'_0| < |u'_0v_0v_1 u'_j \cdots v_0v_1 u'_{i+1}|
  \end{equation}
  since $|w'_0|\leq|u'_0|$ by~(\ref{eq:10}). Let
  \[
    f_{11} = v_1 w_2 u'_0 v_0v_1 u'_j \cdots
             v_0v_1 u'_{i+1} \bar v v_0 \ .
  \]
  Then $|w|<|f_{11}|$ by~\eqref{eq:9}, and hence,
  $f_{11}$ is bordered. Let $h_{11}=v_1g_{11}v_0$
  be the shortest border of $f_{11}$.
  Recall, that $w_2\neq\eps$ and either $\SUF{az}{v_1w_2}$
  or $\SUF{v_1w_2}{az}$. If $|v_1w_2|<|az|$ then $v_1$ necessarily
  occurs in $z$, and hence, it overlaps with $v_0$ (since
  $\PRE{bz}{v_0v_1}$); a~contradiction. So, we have $\SUF{az}{v_1w_2}$.
  Surely, $|h_{11}|<|v_1w_2|$ (and so $\PRE{h_{11}}{v_1 w_2}$)
  for otherwise $az$ occurs in $u$ which contradicts our assumption
  that $z$ is of maximum length. Let $w_2=g_{11}v_0w_5$.
  Note, that $|v_0w_5|\neq|az|$ since $az$ and~$v_0$ begin with
  different letters. We have $|az|<|v_0w_5|$ since otherwise $v_0$
  occurs in $z$, and hence, overlaps with $v_1$ which
  is a~contradiction. Consider,
  \[
    f_{12} = v_0w_5 u'_0 v_0v_1 u'_j \cdots
             v_0v_1 u'_{i+1} \bar v v_0v_1 \ .
  \]
  If $f_{12}$ is unbordered, then $|u|<|w|-1$ since
  $|f_{12}|\leq|w|$ and
  \[
    |u| = |f_{12}| - |v_0 w_5| + |t'|
  \]
  and $|az|<|v_0 w_5|$ and $|t'|\leq|z_0|\leq|z|<|bz|<|v_0w_5|$.
  Assume, $f_{12}$ is bordered. Then
  $f_{12}$ has a shortest border
  $h_{12}=g_{12}v_0v_1$ with $|az|<|h_{12}|$, for otherwise
  $az$ occurs in $u$. Let $v_0w_5=g_{12}v_0v_1w_6$. But, now
  \[
    w = w'_0 \bar v v_0v_1 g_{12} v_0v_1 w_6
  \]
  where $\SUF{v_0v_1w_6}{w_2}$,
  contradicting our assumption that $v_0v_1$ does not occur
  in $w_2$.

  If $i>j$ then
  \[
    w = w'_0 v_0v_1 w'_i \cdots
             v_0v_1 w'_{j+1}\bar v v_0v_1 w_2
    \qquad\text{and}\qquad
    u = u'_0 \bar v v_0v_1 t'
  \]
  and $|w|\geq|u|-|t'|+|v_0v_1|$. We have $|u|<|w|-1$
  since $|t'|\leq|z_0|<|v_0v_1|-1$ by~(\ref{eq:12}).

  Assume $i=j$. Then
  \[
    w = w'_0 \bar v v_0v_1 w_2 \qquad\text{and}\qquad
    u = u'_0 \bar v v_0v_1 t' \ .
  \]
  Consider
  \[
    f' = v_1 w_2 u'_0 \bar v v_0 \ .
  \]
  If $f'$ is bordered, then it has a~shortest border
  $h'=v_1g'v_0$.
  
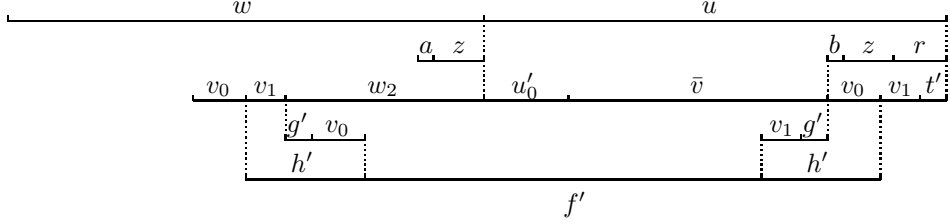
\begin{center}
  \begin{picture}(355,85)(0,0)
    \put(85,78){\makebox{$w$}}
    \put(262.5,78){\makebox{$u$}}
    \drawline(0,77)(0,75)(180,75)(180,77)
    \drawline(180,77)(180,75)(355,75)(355,77)

    \put(155.5,63){\makebox{$a$}}
    \put(168,63){\makebox{$z$}}
    \drawline(155,62)(155,60)(161,60)(161,62)(161,60)
         (180,60)(180,62)
    \dottedline{2}(180,62)(180,75)

    \put(310.6,63){\makebox{$b$}}
    \put(323,63){\makebox{$z$}}
    \drawline(310,62)(310,60)(316,60)(316,62)(316,60)
         (335,60)(335,62)
    \put(342.5,63){\makebox{$r$}}
    \drawline(335,62)(335,60)(355,60)(355,62)
    \dottedline{2}(355,62)(355,75)

    \put(190.5,48){\makebox{$u'_0$}}
    \put(258,48){\makebox{$\bar v$}}
    \drawline(180,47)(180,45)
        (212,45)(212,47)(212,45)
        (310,45)(310,47)
    \dottedline{2}(180,47)(180,60)

    \put(315,48){\makebox{$v_0$}}
    \put(333,48){\makebox{$v_1$}}
    \drawline(310,47)(310,45)
        (330,45)(330,47)(330,45)
        (345,45)(345,47)
    \dottedline{2}(310,47)(310,60)

    \put(347,48){\makebox{$t'$}}
    \drawline(345,47)(345,45)
         (355,45)(355,47)
    \dottedline{2}(355,47)(355,60)

    \put(75,48){\makebox{$v_0$}}
    \put(93,48){\makebox{$v_1$}}
    \put(136,48){\makebox{$w_2$}}
    \drawline(70,47)(70,45)
        (90,45)(90,47)(90,45)
        (105,45)(105,47)(105,45)
        (180,45)(180,47)

    \put(105.5,33){\makebox{$g'$}}
    \put(120,33){\makebox{$v_0$}}
    \drawline(105,32)(105,30)
        (115,30)(115,32)(115,30)
        (135,30)(135,32)
    \dottedline{2}(105,32)(105,45)

    \put(288,33){\makebox{$v_1$}}
    \put(300.5,33){\makebox{$g'$}}
    \drawline(285,32)(285,30)
        (300,30)(300,32)(300,30)
        (310,30)(310,32)
    \dottedline{2}(310,32)(310,45)

    \put(107,18){\makebox{$h'$}}
    \put(302,18){\makebox{$h'$}}
    \drawline(90,17)(90,15)
        (135,15)(135,17)(135,15)
        (285,15)(285,17)(285,15)
        (330,15)(330,17)
    \dottedline{2}(90,17)(90,45)
    \dottedline{2}(135,17)(135,30)
    \dottedline{2}(285,17)(285,30)
    \dottedline{2}(330,17)(330,45)

    \put(210,3){\makebox{$f'$}}
  \end{picture}
  \end{center}
  Recall, that $w_2\neq\eps$ and either $\SUF{az}{v_1w_2}$
  or $\SUF{v_1w_2}{az}$. If $|v_1w_2|<|az|$ then $v_1$ occurs
  in $z$, and hence, overlaps with $v_0$ since $\PRE{bz}{v_0v_1}$;
  a~contradiction. So, we have $\SUF{az}{v_1w_2}$.
  Surely, $|h'|<|v_1w_2|$ for otherwise $az$ occurs in $u$ which
  contradicts our assumption. Let $w_2=g'v_0w_4$.
  Note, that $|v_0w_4|\neq|az|$ since $az$ and $v_0$ begin with
  different letters. We have $|az|<|v_0w_4|$ since otherwise $v_0$
  occurs in $z$, and hence, overlaps with $v_1$ which
  is a~contradiction. Consider now,
  \[
    f'' = v_0w_4 u'_0 \bar v v_0v_1 \ .
  \]
  If $f''$ is unbordered, then it easily follows that
  $|u|<|w|-1$ since we have $|t'|<|az|$ and $|az|<|v_0w_4|$.
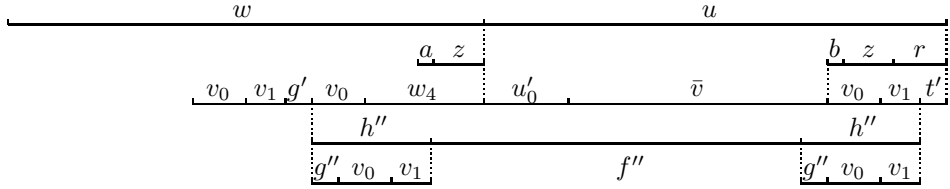
  \begin{center}
  \begin{picture}(355,70)(0,0)
    \put(85,63){\makebox{$w$}}
    \put(262.5,63){\makebox{$u$}}
    \drawline(0,62)(0,60)(180,60)(180,62)
    \drawline(180,62)(180,60)(355,60)(355,62)

    \put(155.5,48){\makebox{$a$}}
    \put(168,48){\makebox{$z$}}
    \drawline(155,47)(155,45)(161,45)(161,47)(161,45)
         (180,45)(180,47)
    \dottedline{2}(180,47)(180,60)

    \put(310.6,48){\makebox{$b$}}
    \put(323,48){\makebox{$z$}}
    \drawline(310,47)(310,45)(316,45)(316,47)(316,45)
         (335,45)(335,47)
    \put(342.5,48){\makebox{$r$}}
    \drawline(335,47)(335,45)(355,45)(355,47)
    \dottedline{2}(355,47)(355,60)

    \put(190.5,33){\makebox{$u'_0$}}
    \put(258,33){\makebox{$\bar v$}}
    \drawline(180,32)(180,30)
        (212,30)(212,32)(212,30)
        (310,30)(310,32)
    \dottedline{2}(180,32)(180,45)

    \put(315,33){\makebox{$v_0$}}
    \put(333,33){\makebox{$v_1$}}
    \drawline(310,32)(310,30)
        (330,30)(330,32)(330,30)
        (345,30)(345,32)
    \dottedline{2}(310,32)(310,45)

    \put(347,33){\makebox{$t'$}}
    \drawline(345,32)(345,30)
         (355,30)(355,32)
    \dottedline{2}(355,32)(355,45)

    \put(75,33){\makebox{$v_0$}}
    \put(93,33){\makebox{$v_1$}}
    \put(105.5,33){\makebox{$g'$}}
    \put(120,33){\makebox{$v_0$}}
    \put(151,33){\makebox{$w_4$}}
    \drawline(70,32)(70,30)
        (90,30)(90,32)(90,30)
        (105,30)(105,32)(105,30)
        (115,30)(115,32)(115,30)
        (135,30)(135,32)(135,30)
        (180,30)(180,32)

    \put(133,18){\makebox{$h''$}}
    \put(318,18){\makebox{$h''$}}
    \drawline(115,17)(115,15)
        (160,15)(160,17)(160,15)
        (300,15)(300,17)(300,15)
        (345,15)(345,17)
    \dottedline{2}(115,17)(115,30)
    \dottedline{2}(345,17)(345,30)

    \put(230,3){\makebox{$f''$}}

    \put(115.5,3){\makebox{$g''$}}
    \put(130,3){\makebox{$v_0$}}
    \put(148,3){\makebox{$v_1$}}
    \drawline(115,2)(115,0)
        (125,0)(125,2)(125,0)
        (145,0)(145,2)(145,0)
        (160,0)(160,2)
    \dottedline{2}(115,2)(115,15)
    \dottedline{2}(160,2)(160,15)

    \put(300.5,3){\makebox{$g''$}}
    \put(315,3){\makebox{$v_0$}}
    \put(333,3){\makebox{$v_1$}}
    \drawline(300,2)(300,0)
        (310,0)(310,2)(310,0)
        (330,0)(330,2)(330,0)
        (345,0)(345,2)
    \dottedline{2}(300,2)(300,15)
    \dottedline{2}(345,2)(345,15)
  \end{picture}
  \end{center}
  If $f''$ is bordered, then it has a shortest border
  \hbox{$h''=g''v_0v_1$} with $|az|<|h''|$, for otherwise
  $az$ occurs in $u$. Let $v_0w_4=g''v_0v_1w_5$. But, now
  \[
    w = w'_0 \bar v v_0v_1 g' g'' v_0v_1 w_5
  \]
  which contradicts our assumption that
  \hbox{$w = w'_0 \bar v v_0v_1 w_2$}
  and $v_0v_1$ does not occur in~$w_2$.

  If $f'$ is unbordered, then $|f'|\leq|w|$, and hence,
  $|w'_0|\geq|u'_0|$.
  But, we also have $|w'_0|\leq|u'_0|$; see~\eqref{eq:10}.
  That implies $|w'_0|=|u'_0|$. Moreover, the factors $w_0$
  and $bzv'$ have both nonoverlaping occurrences in $u'_0v_0v_1$
  by~\eqref{eq:10}. Therefore, $w'_0=u'_0$. Now,
  \[
    w = xa w_7 \qquad\text{and}\qquad u = xb t''
  \]
  where $\PRE{w'_0 \bar v v_0v_1}{x}$
  and $a, b\in A$ and $a\neq b$ and $\SUF{w_7}{w_2}$
  and $\SUF{t''}{t'}$. We have that $xb$ occurs in~$w$
  by Theorem~\ref{thm:1}. Since $xb$ is not a prefix of~$w$
  and $v_0v_1$ does not overlap with itself,
  we have $|xb|+|v_0v_1|\leq|w|$.
  From $|t'|\leq|z_0|<|v_0v_1|-1$ we get
  $|u| < |w| - 1$ and the claim follows.
\end{proof}

Note, that the bound $|u|<|w|-1$ on the length
of a nontrivial Duval extension $wu$ of~$w$
is tight, as the following example shows.
\begin{example}
  Let $w=a^n ba^{n+m}bb$ and $u=a^{n+m}ba^n$ with $n, m\geq 1$.
  Then
  \[
    w.u = a^n ba^{n+m}bb.a^{n+m}ba^n
  \]
  is a nontrivial Duval extension of $w$ and $|u|=|w|-2$.
\end{example}

In general, Duval~\cite{Duval:82} proved that
we have $\per(w)=\mu(w)$, for any word $w$,
if \hbox{$|w|\geq 4\mu(w)-6$}.
Duval also noted that already $|w|\geq 3\mu(w)$ implies
$\per(w)=\mu(w)$, provided his conjecture holds.
Corollary~\ref{cor:1} follows from Theorem~\ref{thm:main}.
\setcounter{EGAL}{\value{theorem}}
\setcounter{theorem}{\value{maincorollary}}
\setcounter{EGAL:sec}{\value{section}}
\setcounter{section}{\value{secintro}}
\begin{corollary}
  If $|w|\geq 3\mu(w)-2$ then \hbox{$\per(w)=\mu(w)$}.
\end{corollary}
\setcounter{section}{\value{EGAL:sec}}
\setcounter{theorem}{\value{EGAL}}
However, this bound is unlikely to be tight. The best example
for a large bound that we could find is taken from~\cite{AssPou:79}.
\begin{example}
  Let
  \[
    w = a^n ba^{n+1} ba^n ba^{n+2} ba^n ba^{n+1} ba^n \ .
  \]
  We have $|w|=7n+10$ and $\mu(w)=3n+6$ and $\per(w)=4n+7$.
\end{example}
So, we have that the precise bound for the length of a word
that implies $\per(w)=\mu(w)$ is larger than $7/3\mu(w)-4$
and not larger than $3\mu(w)-2$. The characterization of the
precise bound of the length of a word as a function of its
longest unbordered factor is still an open problem.

\section{Conclusions}\label{sec:concl}
In this paper we have given a confirmative answer
to a long standing conjecture \cite{Duval:82}
by proving that a Duval extension $wu$ of~$w$
longer than \hbox{$2|w|-2$} is trivial. This
bound is tight and also gives a new bound
on the relation between
the length of an arbitrary word $w$ and its
longest unbordered factors $\mu(w)$, namely
that $|w|\geq3\mu(w)-2$ implies $\per(w)=\mu(w)$
as conjectured (more weakly) in~\cite{AssPou:79}.
We believe that the precise bound can
be achieved with methods similar to those
presented in this paper.

\bibliographystyle{plain}
\bibliography{Bibo}

\end{document}